\documentclass[reprint,aps,prl,twocolumn,showkeys,showpacs,floatfix]{revtex4}
\usepackage{dcolumn}
\usepackage{bm}
\usepackage{graphicx}
\usepackage{bm}

\begin{document}

\title{Evolution of Magnetic Glass on Partial Crystallization of a Bulk Metallic Glass:  Tb$_{36}$Sm$_{20}$Al$_{24}$Co$_{20}$}

\author{Archana Lakhani*}
\affiliation{UGC-DAE Consortium for Scientific Research, University Campus, Khandwa Road, Indore-452001,India}

\author{A. Banerjee }
\affiliation{UGC-DAE Consortium for Scientific Research, University Campus, Khandwa Road, Indore-452001,India}

\author{P. Chaddah}
\affiliation{UGC-DAE Consortium for Scientific Research, University Campus, Khandwa Road, Indore-452001,India}

\author{J.Q. Wang }
\affiliation{Institute of Physics, Chinese Academy of Sciences, Beijing 100080, China}

\author{W.H Wang}
\affiliation{Institute of Physics, Chinese Academy of Sciences, Beijing 100080, China}

\author{*archnalakhani@csr.res.in}

\begin{abstract}
\textbf{Abstract}. A comparative study on as cast and annealed rare earth bulk metallic glass (BMG) with composition Tb$_{36}$Sm$_{20}$Al$_{24}$Co$_{20}$ has been carried out by magnetization measurements. The as cast amorphous sample shows non-ergodic magnetization but does not show the behavior expected from magnetic glass. After annealing, the partially crystallized BMG shows this magnetic glass behavior. This is confirmed by the established measurement protocol of cooling and heating in unequal fields (CHUF).
\end{abstract}



\maketitle

\section{\textbf{INTRODUCTION}}
Structural glasses are disordered materials which lack crystalline periodicity and are mechanically solids. Most metals crystallize on cooling, arranging their atoms into a regular pattern called lattice, but if crystallization does not occur and the atoms settle in a random arrangement, the final form is a metallic glass\cite{Greer}.  The critical cooling rate is much slower for BMG's, enabling metglass samples of larger dimensions\cite {Inoue,Luo}.

A similar kind of arrested kinetics of the first order magnetic transition from ferromagnetic (FM) to antiferromagnetic (AFM) or vice versa has been observed in half doped Manganites and in various intermetallic alloys like doped CeFe$_2$, Co doped Mn$_2$Sb, FeRh and ferromagnetic shape memory alloys in recent years. These materials have been examined thoroughly by various protocols including the novel protocol CHUF conceived and developed to identify the magnetic glasses and its low temperature equilibrium state \cite{Banerjee,Banerjee1}. The commonality between the structural and magnetic glass is the arrested kinetics resulting in the persistence of high temperature state as the metastable state at low temperature.

In this manuscript, we highlight on the possibility of a rare-earth bulk metallic glass with composition Tb$_{36}$Sm$_{20}$Al$_{24}$Co$_{20}$ to be a magnetic glass. We have compared the magnetization behavior of as cast and annealed specimen in order to find the correlation between structural glass and magnetic glass.

\section{\textbf{EXPERIMENTAL}}
The as cast BMG-Tb$_{36}$Sm$_{20}$Al$_{24}$Co$_{20}$ sample is prepared by arc melting the pure elements in an argon atmosphere and then suck - cast into a Cu mold to get a cylindrical rod of 2 mm diameter. A small piece of this sample was crystallized by annealing at 450$^\circ$C. The amorphous and crystalline nature of the sample is confirmed by X-ray diffraction using Bruker (D8 Advance) X-ray diffractometer with Cu K$\alpha$ radiations. Magnetization measurements were carried out using a commercial 14T-VSM PPMS.

\section{\textbf{RESULTS AND DISCUSSIONS}}
The reported glass transition temperature (T$_g$), crystallization temperature (T$_x$), and the melting temperature (T$_m$) of the BMG Tb$_{36}$Sm$_{20}$Al$_{24}$Co$_{20}$ is 309$^\circ$C, 383$^\circ$C and 659$^\circ$C respectively \cite{Luo}. BMGs are partially crystallized by isothermal annealing at a temperature in the supercooled liquid region or by fast annealing just above crystallization temperature \cite{Inoue}. This is known as partial devitrification of BMGs which gives rise to the formation of composites or multi-phases. The magnetic behavior is complex especially in RE based BMGs when multiphases are present after tailoring them by different heat treatments. In order to understand whether the inhomogeneous structure arising from multi-phases can hinder the critical dynamics, we have performed the Zero Field cooled (ZFC) and Field cooled magnetization (FC) measurements at various fields on as cast as well as annealed sample.

\

\begin{figure}[!h]
\begin{centering}
\includegraphics[width=0.5\columnwidth]{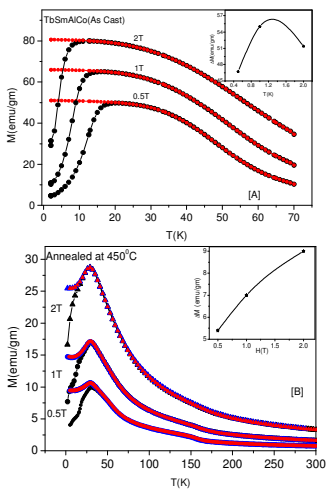}
\par\end{centering}
\caption{(A-B): ZFC and FC Magnetization w.r.t Temperature at 0.5T, 1 and 2T on as cast and annealed Tb$_{36}$Sm$_{20}$Al$_{24}$Co$_{20}$.} \label{fig: Fig1}
\end{figure}

Figure 1(a) shows ZFC and FC behavior of the M(T) measurements at 0.5T, 1T and 2T on as cast sample. The bifurcation between the ZFC and FC curves is an indication of non-ergodicity. This temperature where the bifurcation between two curves begin,  decreases with increase in magnetic field suggesting the irreversibility process begins at lower temperature as field rises. The difference in ZFC and FC magnetization at low temperature is defined as the thermomagnetic irreversibility ($\Delta$M =$ M_{ZFC@5~K} - M_{FC@5~K}$) which increases upto 1T and then decreases thereafter at 2T for as cast sample as shown in the inset of figure1(A). Similar type of non-ergodic behavior and  $\Delta$M has been observed in spin glasses.
Figure 1(B) shows the ZFC and FC magnetization behavior of the crystalline counter part at 0.5, 1 and 2T   which shows an antiferromagnetic transition (T$_C$) at $\sim$ 30K and it decreases to 29K at 2T.  This FM-AFM transition is not seen in the as cast sample. Inset of figure 1(B) shows the thermo-magnetic irreversibility behavior which increases with field in contrary to the as cast sample.  $\Delta$M rising with field has been observed in magnetic glasses mentioned above.

To validate the kinetically arrested state in the crystalline counter part we have performed the well known protocol CHUF on as cast as well as on the partially crystallized sample as shown in figure 2. Figure 2 (A and B) shows the temperature dependent magnetization at 1T and 2T after cooling in various fields on the as cast sample. Magnetization measurements at 1T after cooling in 0, 0.5,1, 2, 3, and 10T demonstrates  the non-ergodic behavior of as cast sample below 14K, which reduces to 13K on increasing the warming field to 2T. On the other hand M(T) behavior on annealed sample shows two transitions when cooling field (H$_C$) is higher than the warming field (H$_W$) while only one transition when H$_{C} \leq  H_W$ as shown in figure 2(C and D). The sudden fall in magnetization at low temperatures for H$_{C} > H_W$ signifies the devitrification of the magnetic glassy state and we note that for H$_{C} > H_W$ we have non-ergodic behavior below $\sim $ 15K  where as for H$_{C} < H_W$, it is observed at a higher temperature at $\sim $ 20K.

\begin{figure} [!h]
\begin{centering}
\includegraphics[width=0.85\columnwidth]{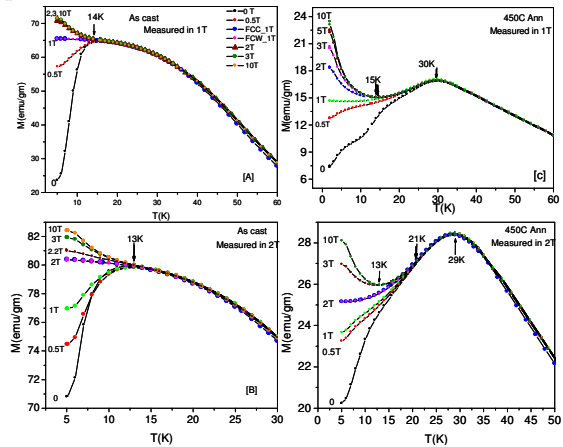}
\par\end{centering}
\caption{(A-D): M(T) at 1 and 2T after cooling in various fields marked on as cast and annealed sample of  Tb$_{36}$Sm$_{20}$Al$_{24}$Co$_{20}$.} \label{fig: Fig2}
\end{figure}

Hence, our measurements show that the as cast amorphous BMG sample does not become a magnetic glass whereas the annealed (partially crystallized) sample resembles a magnetic glass.

\end{document}